**Effect of nano Co reinforcements on the structure of the Sn-3.0Ag-0.5Cu solder in liquid and after reflow solid states**

Andriy Yakymovych[1,2,*], Stepan Mudry[2], Ihor Shtablavyi[2], Herbert Ipser[1]

[1]Department of Inorganic Chemistry (Materials Chemistry), University of Vienna, Währinger Str. 42, 1090 Vienna, Austria
[2]Department of Metal Physics, Ivan Franko National University of Lviv, Kyrylo i Mephodiy str. 8, 79005 Lviv, Ukraine

Corresponding author:
Andriy Yakymovych
Department of Inorganic Chemistry (Materials Chemistry), University of Vienna, Währinger Str. 42, 1090 Vienna, Austria
Department of Metal Physics, Ivan Franko National University of Lviv, Kyrylo i Mephodiy str. 8, 79005 Lviv, Ukraine

e-mail: yakymovych@univie.ac.at

Sn-Ag-Cu (SAC) alloys are commonly recognized as lead-free solders employed in the electronics industry. However, some disadvantages in mechanical properties and their higher melting temperatures compared to Pb-Sn solders prompt new research relating to reinforcement of existing SAC solders. One of the ways to reinforce these solder materials is the formation of composites with nanoparticles as filler materials. Accordingly, this study presents structural features of nanocomposite $(Sn\text{-}3.0Ag\text{-}0.5Cu)_{100-x}(nanoCo)_x$ solders with up to 0.8 wt.% nano Co. The effect of nano-sized Co particles was investigated by means of differential thermal analysis (DTA), X-ray diffraction (XRD) in both liquid and solid states, and scanning electron microscopy (SEM). The experimental data of DTA are compared with available literature data for bulk Sn-3.0Ag-0.5Cu alloy to check the capability of minor nano-inclusions to decrease the melting temperature of the SAC solder. The combination of structural data in liquid and solid states provides important information about the structural transformations of liquid Sn-3.0Ag-0.5Cu alloys caused by minor Co additions and the phase formation during crystallization. Furthermore, scanning electron microscopy has shown the mutual substitution of Co and Cu atoms in the $Cu_6Sn_5$ and $CoSn_3$ phases, respectively.

## 1. Introduction

A multitude of investigations has been performed to study different physico-chemical properties of lead-free Sn-Ag-Cu (SAC) solder alloys and to analyze their structure in the solid and liquid state [1-7]. However, due to their higher melting temperatures compared to Pb-Sn solders and the formation of brittle intermetallic compound layers, a number of recent studies addressed the possibility to improve the characteristics of SAC solders by adding of fourth element in the bulk or nano-sized form, such as Co, Ni, Fe or Al [8-12]. For instance, it was shown by Anderson et al. [8] that the shear strength of joints made from Sn-3.0Ag-0.5Cu solder was the lowest. To improve mechanical properties, bulk Co and Fe were added in Refs. [8, 9]. Additionally, a previous study of the effect of Co additions in a solute-enriched Sn-3.6Ag-1.0Cu-0.45Co alloy indicated that this transition metal acts as a solidification catalyst reducing undercooling and increasing the nucleation site density for $Cu_6Sn_5$ [13]. In Ref. [10], the growth of the brittle Cu-Sn IMCs was suppressed by minor additions of Al up to 1 wt.%. In this case, the reduction in the growth rate of brittle Cu-Sn IMCs was attributed to the initial formation and evolution of Al-Cu phase layers. Additions of Ni into the lead-free SAC solder promoted the formation of new $(Cu,Ni)_6Sn_5$ particles refining the microstructure of solder joints and caused a reduction of undercooling and a decrease of the solidus temperature [11, 12].

Many other studies reported the refinement of solder microstructure and the improvement of mechanical properties resulting from the addition of nanoparticles into SAC solders [14-22]. Moreover, metallic nano-size particles, such as Al, Fe, Zn, Ni, and Co, can react with the elements Sn, Ag, Cu in the molten solder during reflow, and the particles can be dissolved in the solder matrix or can participate in forming the interfacial IMC layer [23]. In general, either metal nanoparticles were added directly into the SAC solder paste, or a flux was added to a pre-mixed (SAC + nanoparticles) powder in preparing a solder paste [15-18]. Therefore, these studies are mainly focused on thermal, mechanical properties and the microstructure of solder joints.

To the best of our knowledge, few papers are related to the introduction of oxide or ceramics nano-particles into the metal matrix, namely by mechanically mixing of nano-additions with SAC powder using either ball milling technique [14] or cold-compacting afterwards [19]. Other methods relate to introducing of nanoparticles into the SAC alloy melt combined with mechanical stirring [20, 21] or via the accumulative roll-bonding (ARB) process [22]. However, the studies mentioned above relate to introducing of non-reacting nano-inclusions into the SAC solders, such as single-walled carbon nanotubes, $Al_2O_3$, SiC or $CeO_2$, which do not react with the metal matrix.

A key point in the present work is to examine the impact of nano Co inclusions on the melting temperature and structural features of the liquid Sn-3.0Ag-0.5Cu solder. Furthermore, a structure analysis of as-solidified (Sn-3.0Ag-0.5Cu)-Co alloys indicates changes in the microstructure of the Sn-3.0Ag-0.5Cu solder in the solid state caused by minor Co impurities.

## 2. Materials and methods

The composite $(SAC305)_{100-x}(nanoCo)_x$ alloys were prepared by mixing Sn-3.0Ag-0.5Cu (composition in mass %; from now on SAC305) alloy in powder form (average size 31 μm, SAC305 powder, Kester, U.S.A.) with 0.2, 0.5 and 0.8 wt. % Co nanoparticles (average size 28 nm, 99.8%, IOLITEC GmbH, Heilbronn, Germany). Mechanical dispersion of the nanoparticles in the solder powder was achieved by mechanically mixing, maintaining a frequency of 700 rpm (11.7 $s^{-1}$) at room temperature for approximately 30 min using a Retsch mixer (Retsch MM301). After that the $(SAC305)_{100-x}(nanoCo)_x$ powders were pressed into pellet form (about 13 mm diameter and about 5 mm height). All operations with Co nanoparticles were performed in a glove box (M.Braun, LabMaster 130) under an atmosphere of purified Ar inside ($O_2$ and $H_2O$ < 5ppm each).

Differential thermal analysis (DTA) was performed to identify phase transformations and the corresponding temperatures using a Netzsch DTA 404F. The Pt/Pt10%Rh thermocouples were calibrated at the melting points of pure Sn, Sb and In. For each sample two heating and cooling cycles were performed in open alumina crucibles under a constant argon flow of 50 mL·min$^{-1}$ and with a heating/cooling rate of 5 K·min$^{-1}$.

High-temperature XRD measurements were carried out by means of a high-temperature X-ray diffractometer with a special attachment, which allows investigation of the solid and liquid samples at different temperatures. Cu$K\alpha$ radiation monochromatized with a LiF single crystal and Bragg–Brentano focusing geometry in the $2\theta$ range from $10^0$ up to $120^0$ were used. Intensity curves were corrected for polarization and incoherent scattering [24]. After this procedure they were normalized to electron units by the Krogh-Moe method [25].The structure factors (SF), the atomic pair distribution functions, and functions of the radial distribution of atoms were calculated from the corrected and normalized intensity curves. The scattering intensities were measured with an accuracy of better than 2%, and the temperature was determined and maintained with an accuracy of ±2 K.

Powder XRD measurements were done on a Bruker D8 diffractometer at ambient temperature using Ni filtered $CuK_{\alpha}$ radiation (accelerating voltage 40 kV, electron current 40 mA). The diffractometer operates in the $\theta/2\theta$ mode. The powder was fixed with petroleum jelly on a single crystal silicon sample carrier which was rotated during the measurement. The detection unit was a Lynxeye strip detector. Indexing of the phases was supported by the Inorganic Crystal Structural Database (ICSD). Rietveld refinement of the XRD patterns was done with the Topas3® software provided by Bruker AXS.

The scanning electron microscope Zeiss Supra 55 VP (SEM) was used for metallographic investigations. The excitation energy of the electron beam was 15-20 kV; backscattered electrons (BSE) were detected in order to visualize the surfaces of the samples. The chemical analyses of the sample phases were performed using the energy dispersive X-ray (EDX) technique with four characteristic spectral lines of Co, Cu (K-line) and Ag, Sn (L-line). Pure Co was also used for energy calibration of the EDX detector signal. An acceleration voltage of 20 kV was applied. The standard deviation for the chemical compositions obtained from EDX is about ±1 at%.

### 3. Results and discussion

The results of DTA combined with XRD and SEM analysis of samples after reflow were used to identify the structure transitions of SAC305 solder reinforced by nano-sized Co. Several composite $(SAC305)_{100-x}(nanoCo)_x$ alloys were studied by DTA to determine phase transitions and reaction temperatures (see Table 1).

**Table 1**

Summary of compositions as well as melting temperatures ($T_l$).

| Composition, wt.% | | $T_l$, K |
|---|---|---|
| SAC305 | Co | |
| 100 | 0 | 489.3±0.5 |
| 99.8 | 0.2 | 488.9*<br>490.3 |
| 99.5 | 0.5 | 489.0*<br>490.4 |
| 99.2 | 0.8 | 489.3*<br>490.8 |

* corresponds to the nanocomposite $(SAC305)_{100-x}(nanoCo)_x$ alloys (first heating).

As mentioned in the experimental section, for each sample two heating and cooling cycles were performed at a heating rate of 5 K $min^{-1}$. We expected that during first heating in the temperature region above the melting temperature of SAC305 the nano-sized Co particles would dissolve in the liquid SAC 305 alloy. Therefore, the first cooling run and the second heating/cooling cycle of DTA should relate to the quaternary Ag-Co-Cu-Sn system. In this case, the reaction temperatures from first heating correspond to the composite $(SAC305)_{100-x}(nanoCo)_x$ alloys and, beginning with the first cooling, they should correspond to quaternary $(SAC305)_{100-x}(Co)_x$ alloys. The liquidus effects for nanocomposite alloys were evaluated on first heating while, for quaternary alloys, they were evaluated on second heating; all other effects were taken from the cooling curves only. The respective reaction temperatures, evaluated from the present results, are listed in Table 1.

According to the phase diagram of the Co-Sn system [26], the Sn-based subsystem shows a Sn-$CoSn_3$ eutectic with the eutectic point at about 1 wt. % Co and a eutectic temperature of about 502 K, which is roughly 3 K below the melting point of pure Sn ($T_l$ = 505 K). Therefore, it is expected that minor Co additions to the Sn-based near-eutectic SAC 305 alloy should lead to insignificant changes in its melting temperature. Particularly, the DTA analysis of Sn-3.6Ag-1.0Cu and Sn-3.6Ag-1.0Cu-0.45Co alloys performed by Anderson et al. [27] indicated that the onset of the endothermic melting event for each alloy occurs at about 500 K and is essentially unaffected by the Co addition. More detailed calorimetric studies of the Sn-3.6Ag-1.0Cu-0.45Co alloy suggested that this minor Co addition does not increase the liquidus temperature by more than about 3 K [28].

As can be seen from Table 1, in contrast to the practically unaffected melting temperature of composite $(SAC305)_{100-x}(nanoCo)_x$ alloys, the melting temperature of quaternary $(SAC305)_{100-x}(Co)_x$ alloys changes slightly by about 1 K for a Co addition of 0.8 wt.%. Furthermore, the solidification temperature of Co reinforced SAC 305 alloys is higher compared to the master alloy indicating a decrease of the difference between melting and solidification temperatures, i.e. a clear reduction of undercooling (Fig. 1). These results suggest a catalytic action of Co for the nucleation of solidification in this alloy system, in agreement with Ref. [13]. It should be noted that undercooling effects in Ref. [13] reached more than 100 K caused by a cooling rate of about 40 K·$min^{-1}$. It should also be noted, that the solidification of quaternary $(SAC305)_{100-x}(Co)_x$ alloys appears within a singular peak, which shows a comparatively small shoulder on the right hand side compared to pure SAC305 alloy (Fig. 1).

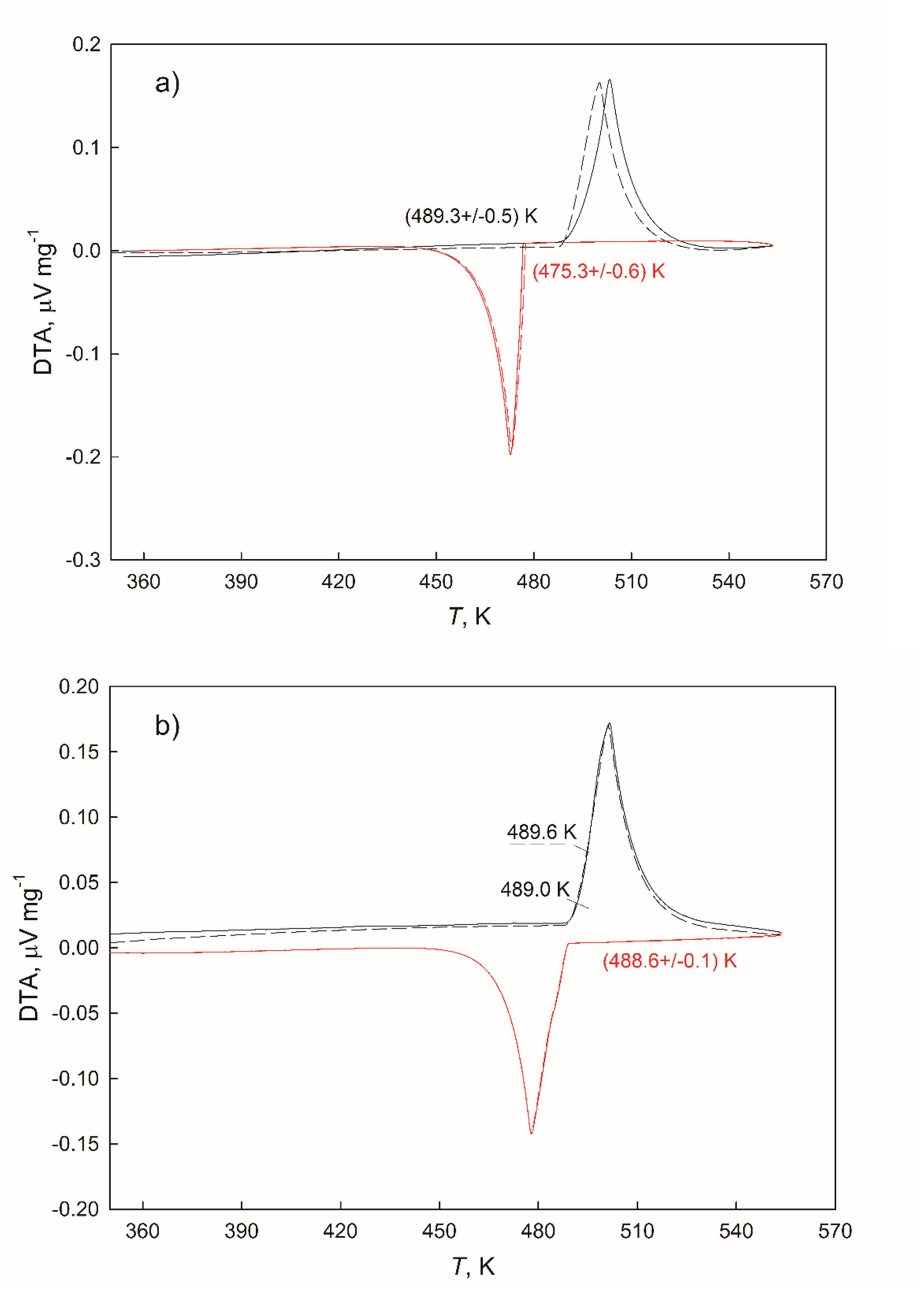

**Fig. 1.** Thermogram sections from DTA measurements of SAC305 (a) and $(SAC305)_{99.5}(nanoCo)_{0.5}$ (b) alloys (black curves correspond to the heating and red curves to the cooling; solid curves corresponds to the first run and dashed lines to the second run).

High-temperature X-ray diffraction of composite $(SAC305)_{100-x}(nanoCo)_x$ alloys was performed at 523 K. This temperature was chosen to discover structure features of the SAC305 solder with minor nano Co additions at a temperature about 30 K above the melting temperature. In fact, nano-sized Co particles should be dissolved in the liquid Sn-based matrix below 523 K. Therefore, the samples are in the homogeneous liquid state at this temperature, and thus the prefix "nano" is deleted for Co in these samples. Fig. 2 shows the structure factors (SFs) for the investigated liquid alloys compared to pure Sn. The right hand side of the principal peak for pure Sn reveals a shoulder interpreted as an evidence of residual covalent bonds in the liquid state.

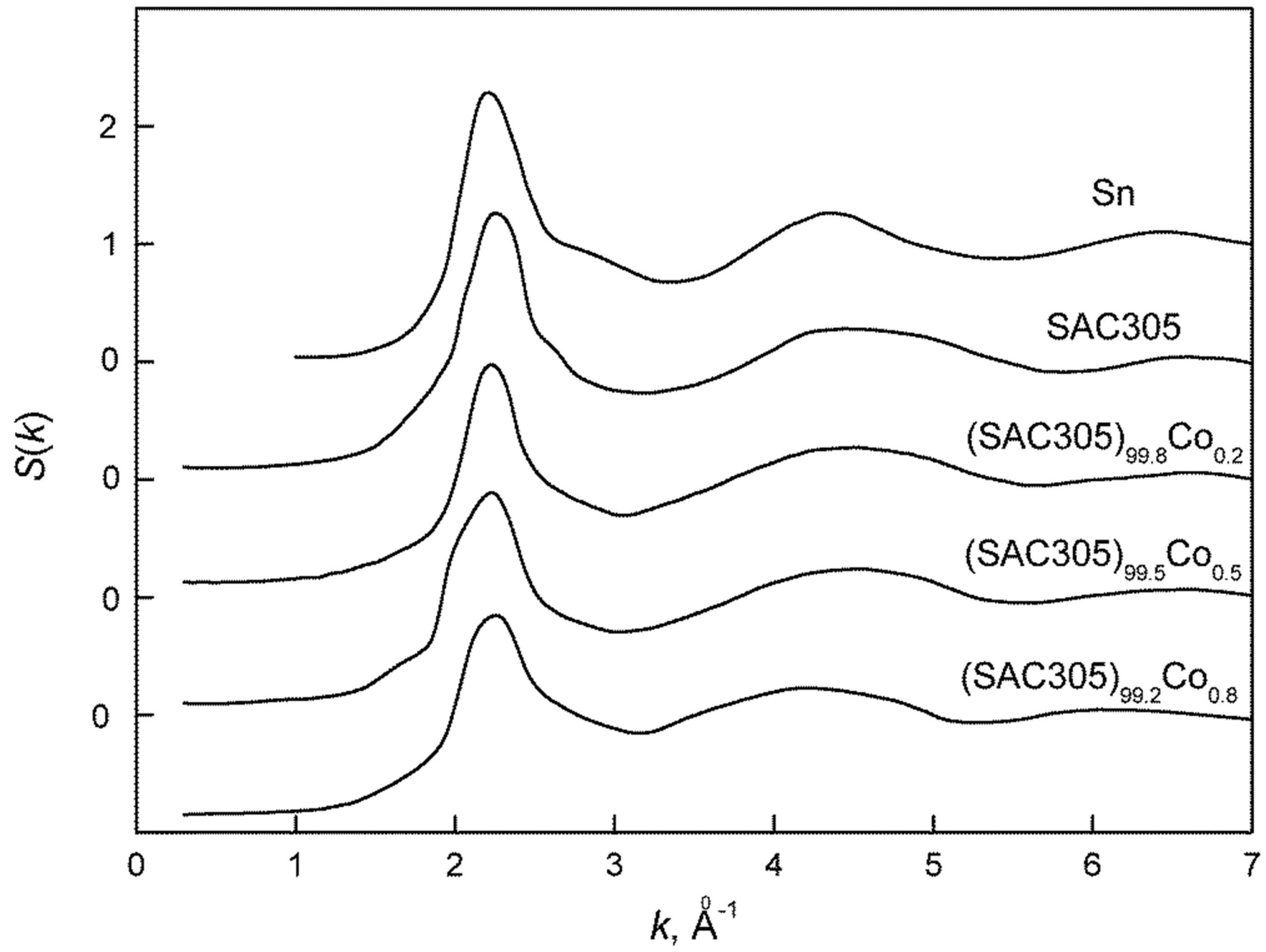

**Fig. 2.** Structure factors of liquid Sn, SAC305 and $(SAC305)_{100-x}(Co)_x$ alloys at 523 K.

As can be seen from Fig. 2, the profile of the SF of pure Sn is insignificantly affected by minor additions of Ag and Cu. Similarly as reported in Ref. [29], Ag and Cu impurities lead to a transformation of the shoulder profile on the right hand side of the principal peak. The addition of 0.2 wt. % Co to the SAC305 alloy leads to a decrease of the principal peak height $S(k_1)$ and to a "smoothing" of the right hand side shoulder. Furthermore, taking into account an increase of the most probable interatomic distances $r_1$ and $r_2$ (Table 2) we suggest that 0.2 wt. % Co promotes the topologic disordering in the atomic arrangement of SAC305 alloy. Moreover, this tendency is much more pronounced for further additions of Co. In addition, the $(SAC305)_{99.5}(Co)_{0.5}$ alloy shows a shoulder on the left hand side of the principal peak. As is can be seen from Fig. 2, the principal peak "absorbs" the right hand shoulder for the alloy with 0.8 wt. %. At the same time the height $S(k_1)$ decreases and the width increases, which corresponds to further disordering of the atomic rearrangement in the liquid SAC305 alloy. Furthermore, a decrease in the $r_2/r_1$ relationship for the $(SAC305)_{99.2}(Co)_{0.8}$ alloy reveals significant structural transformations within the first and second coordination spheres compared to other alloys. In addition, the half-height width of the principal peak was analyzed which gives information required for the estimation of the size of the structural units:

$$L=(2\pi^3)/(2.5^2\cdot\Delta k) \quad (1)$$

where $L$ is the size of the structural units; $\Delta k$ is the half-height width of the principal peak. Initially, Co additions up to 0.5 wt. % lead to any significant changes in the half height width of the principal peak: $\Delta k$ is equal to 0.60 Å$^{-1}$ for the SAC305 alloy; 0.56 Å$^{-1}$ for the $(SAC305)_{99.8}(Co)_{0.2}$ alloy and 0.64 Å$^{-1}$ for the $(SAC305)_{99.5}(Co)_{0.5}$ alloy, respectively. Further increase of Co up to 0.8 wt.% leads to an increase in $\Delta k$=0.92 Å$^{-1}$. The indicated increase of the half height width of the principal peak corresponds to the decrease of the size of the structural units, while the increase of $r_1$ with a decrease of $r_2$ reveals the structural disordering with some changes in interatomic bonding in the SAC305 alloy with 0.8 wt.% Co.

It should be noted that, based on our technique with $CuK_\alpha$ radiation, we could not obtain efficient intensity from the comparatively small amount of Co due to the absorption effect. This allowed us to consider the diffraction pattern as one in which Co atoms are invisible although they affect the structure. In other words, we have diffraction patterns in which there are pores instead of Co atoms which results in an additional increase of the most probable interatomic distance. Therefore, the decrease of the principal peak height could be partly connected with the presence of the Co atoms on the surface layer of the melt. But, an increase of the first most probable interatomic distance by an increase of the Co amount in the liquid $(SAC305)_{100-x}(Co)_x$ alloys also confirm our suggestion about changes of atomic distribution. Such disordering in the frame of the first coordination sphere might be connected with the penetration of Co atoms into Sn clusters or the substitution of other atoms within them, similar to additions of minor amounts of Cu to liquid Sn [29].

**Table 2**
Interatomic distances of the pure Sn and liquid $(SAC305)_{100-x}(Co)_x$ alloys.

| Sample | $r_1$, nm | $r_2$, nm | $r_2/r_1$ |
|---|---|---|---|
| Sn | 3.23 | 6.30 | 1.95 |
| SAC305 | 3.11 | 6.19 | 1.99 |
| $(SAC305)_{99.8}(Co)_{0.2}$ | 3.18 | 6.28 | 1.97 |
| $(SAC305)_{99.5}(Co)_{0.5}$ | 3.21 | 6.34 | 1.97 |
| $(SAC305)_{99.2}(Co)_{0.8}$ | 3.29 | 6.19 | 1.88 |

The microstructure of the solidified samples after structure measurements in the liquid state was studied using ESEM microscope and X-ray diffraction. The results of phase analyses along with BSE images of the investigated alloys can be found in Table 3. According to the obtained results, no residual pure Co was found in samples. Meanwhile, Co atoms substitute Cu atoms in the $Cu_6Sn_5$ compound, which was also observed by Gao et al. [30] studying the morphology and grain growth pattern of intermetallic compounds formed between $Ag_{3.5}Sn_{96.5}$ solder, doped with a small amount of Co, and the Cu substrate. In Ref. [31] it was found that the composition of the $Cu_6Sn_5$ intermetallic in the $(Ag_{3.5}Co_{0.1}Sn_{96.4})$/Cu diffusion couple is $(Cu_{90}Co_{10})_6Sn_5$ on the solder side and $(Cu_{99}Co_1)_6Sn_5$ on the $Cu_3Sn$ side; this shows that Co can substitute Cu atoms in $Cu_6Sn_5$, in agreement with our results presented in Table 3.

It was also found that Cu substitutes Co atoms in the $CoSn_3$ phases. For example, the BSE images of the $(SAC305)_{99.8}Co_{0.2}$ alloy show a uniform distribution of very fine needle-shaped $Ag_3Sn$ phase as well as spherical grains of $(Cu,Co)_6Sn_5$ and $(Co,Cu)Sn_3$. At the same time, neither pure $Cu_6Sn_5$ nor $CoSn_3$ phase was found in the sample. Obviously, the addition of Co atoms into the SAC305 alloy leads to the growth of the $(Co,Cu)Sn_3$ grains, reaching up to 20 μm in diameter, for the $(SAC305)_{99.2}Co_{0.8}$ alloy while the microstructure remains a uniform distribution of very fine other phases in the βSn matrix. The formation of $(Cu,Co)_6Sn_5$ and $(Co,Cu)_3$ phases had been also indicated by the microstructural investigations of solidified $(SAC387)_{100-x}(Co)_x$ samples after calorimetric measurements in Ref. [32].

**Table 3**

SEM-EDX results of $(SAC305)_{100-x}Co_x$ samples after X-ray diffraction analysis in the liquid state.

| Sample | Sn phase | | Ag-Sn phase | | | Cu-Sn phase | | | (Co,Cu)Sn phases | | | | SEM image |
|---|---|---|---|---|---|---|---|---|---|---|---|---|---|
| | | Sn at. % | | Ag at. % | Sn at. % | | Cu at. % | Sn at. % | | Co at. % | Cu at. % | Sn at. % | |
| SAC305 | βSn | 100 | $Ag_3Sn$ | 75 | 25 | $Cu_6Sn_5$ | 53 | 47 | | | | | 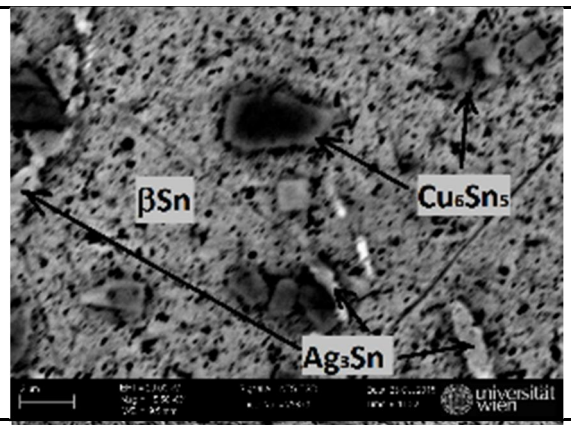 |
| BA1 | βSn | 100 | $Ag_3Sn$ | 74 | 26 | | | | $(Co,Cu)_6Sn_5$<br>$(Co,Cu)Sn_3$ | 1-3<br>17-23 | 47-49<br>2-5 | 48-50<br>74-77 | 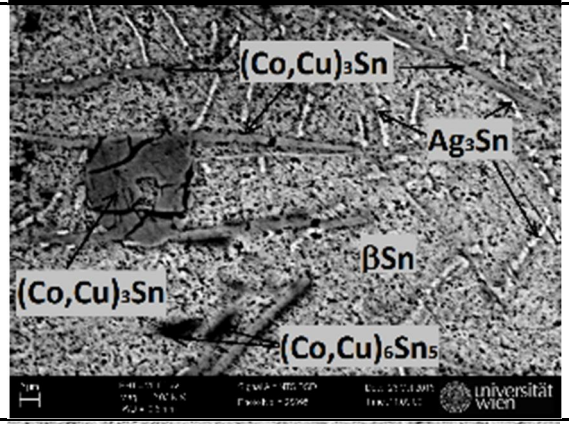 |
| BA2 | βSn | 100 | $Ag_3Sn$ | 75 | 25 | | | | $(Co,Cu)_6Sn_5$<br>$(Co,Cu)Sn_3$ | 1-2<br>19-20 | 48-49<br>3-4 | 49-50<br>76-77 | 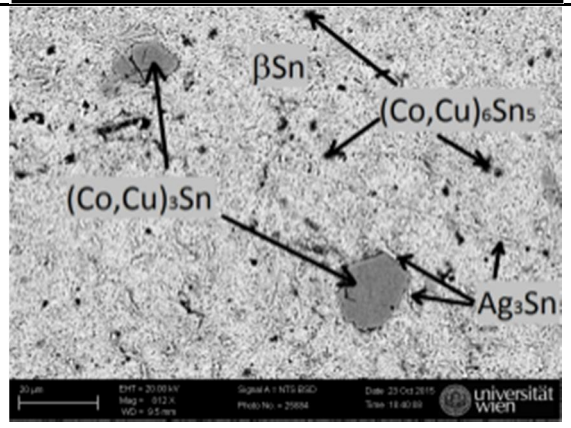 |
| BA3 | βSn | 100 | $Ag_3Sn$ | 74 | 26 | | | | $(Co,Cu)_6Sn_5$<br>$(Co,Cu)Sn_3$ | 1-2<br>19-21 | 48-49<br>2-4 | 49-50<br>76-77 | 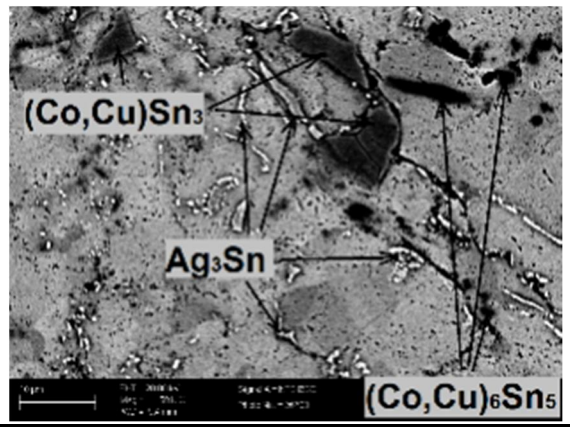 |

BA1 – bulk $(SAC305)_{99.8}Co_{0.2}$ alloy; BA2 – bulk $(SAC305)_{99.5}Co_{0.5}$ alloy; BA3 – bulk $(SAC305)_{99.2}Co_{0.8}$ alloy.

XRD investigations showed the presence of four phases in the investigated samples (Fig. 3) confirming the data in Table 3. It should be noted, that the peaks of binary phases are comparatively weak; therefore, without SEM-EDX analysis it would be difficult to confirm the existence of $Ag_3Sn$, $Cu_6Sn_5$ and $\alpha CoSn_3$ inclusions in the $\beta Sn$ matrix based only on XRD data.

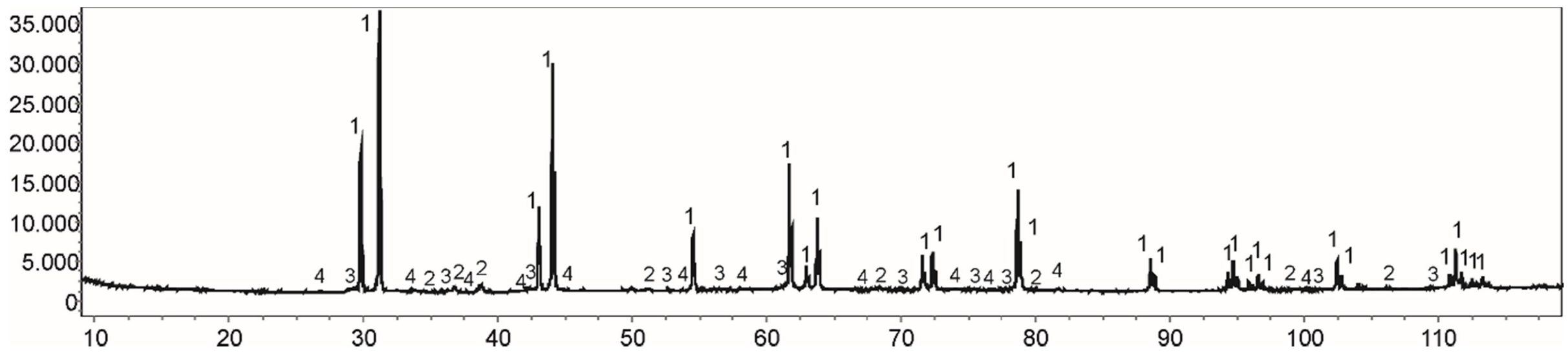

**Fig. 3.** XRD pattern of the $(SAC305)_{99.5}Co_{0.5}$ alloy after X-ray analysis in the liquid state (1 - $\beta Sn$; 2- $Ag_3Sn$; 3 – $Cu_6Sn_5$; 4 – $\alpha CoSn_3$).

Recent investigations of solder joints made from SAC solder paste with minor nano Co additions did not mention any $CoSn_3$ or $(Co,Cu)Sn_3$ phases present inside the joints [15, 16, 31]. Nonetheless, the main topic of these studies was the growth of IMCs at the interface solder/Cu. In contrast, in Ref. [23] it was found that liquid Sn-Ag solder reinforced with up to 2 wt.% Co particles (particle size below 2 μm) reacts with Cu substrate forming $(Co,Cu)_3Sn_2$ IMCs, generally located in the middle of the Sn region. However, no other study could confirm these results. Actually, according to the Co-Sn phase diagram the formation of the $CoSn_3$ phase is expected [26].

## 4. Conclusions

The current work presents a study on SAC305 solder with minor nano-sized Co additions. One of the main open questions for these nanocomposite solders is the dissolution of Co nanoparticles in the molten solder: it is not clear, whether the particles dissolve partially or completely. The presented results indicate complete dissolution of up to 0.8 wt. % of the finely dispersed metallic Co nanoparticles in the SAC305 alloy at temperatures above the melting temperature of this alloy. Furthermore, DTA analysis of composite $(SAC305)_{100-x}(nanoCo)_x$ alloys shows that the melting temperature for the investigated alloys with dissolved Co nanoparticles increases by up to about 1 K by 0.8 wt. %. The microstructure of as-solidified solders contains the phases $Ag_3Sn$, $(Cu,Co)_6Sn_5$, and $(Co,Cu)Sn_3$ homogeneously distributed in the $\beta Sn$ matrix. The minor Co additions lead to a uniform distribution of the structure constituent both in the liquid and solid state; in contrast to the other IMCs, the $(Co,Cu)Sn_3$ phase grows larger in size with increasing of Co content.

**Acknowledgements**

Financial support for this study came from the Austrian Science Fund (FWF) under project Nos. P26304 and P27049. The authors want to thank Dr. Stephan Puchegger from the Department of Physics, University of Vienna, for his help with SEM measurements.